\documentstyle[preprint,tighten,aps,epsfig]{revtex}
\begin{document}
\draft
%
%
%
%
\title{ Transport equations including many-particle correlations for
  an arbitrary quantum system. General formalism}
\author{Jens Fricke}
\address{ Institut f\"{u}r Theoretische Physik der Universit\"{a}t
  G\"{o}ttingen, \\ 
Bunsenstra{\ss}e 9, 37073 G\"{o}ttingen, Germany}
\date{July 4, 1996}
\maketitle
\begin{abstract}
  We present a new method to derive transport equations for
  non-relativistic quantum
  many-particle systems. This method uses an equation-of-motion
  technique and is applicable to interacting fermions and (or) bosons
  in arbitrary time-dependent external fields. Using a
  cluster expansion of the $r$-particle density matrices the infinite
  hierarchy of equations of motion for many-particle expectation
  values is transposed into an equivalent one in terms of
  correlations. This new hierarchy permits a systematic breaking of
  the hierarchy at any order. Diagrams are derived for these transport
  equations. In a second paper the method is tested for exactly
  soluble electron-phonon models in one dimension. 
\end{abstract}
\narrowtext
%
%
%
%
%
%
\section{Introduction}
\label{sec:Intro}
In this work, the first one in a series of two papers on transport
equations, we present a new formalism for the time evolution of
one-particle distribution functions which allows the inclusion of
many-body correlations. In the second paper we derive the transport
equations for an electron-phonon system and compare the numerical
results with the exact solutions of one-dimensional models (polaron
model, Tomonaga-Luttinger model).

Very often non-equilibrium phenomena in quantum many-particle systems
are studied by the use of real-time Green's function techniques
(Martin and Schwinger \cite{MS,Sw}, Kadanoff and Baym \cite{KB},
Keldysh \cite{Ke}, Langreth and Wilkins \cite{LW,La}). For a review
see Ref.\ \cite{Ra}. In the Keldysh formalism the perturbative
expansion uses a time integration over a path in the complex plane. If
only interested in the one-particle distribution function, one usually
applies the generalized Kadanoff-Baym ansatz \cite{Li,Ha,KB}, which
was justified by Schoeller \cite{Sch} as a partial resummation in the
diagrammatic expansion. The transport equation obtained in the Born
approximation is of the form of a non-Markovian Boltzmann equation
with renormalized one-particle propagators, usually taken the same as
in equilibrium. In Ref.~\cite{Sch} initial correlations are also
discussed, but it does not seem to be possible to apply the same
resummation method including the initial correlations. This stems from
the fact that the correlations are not taken into account as
time-dependent quantities.

Motivated by the work of Schoeller, we present a method which enables
the inclusion of the correlations as dynamical quantities. Only
single-time expectation values of operators are considered in contrast
to Green's function techniques. In the course of this work (cf.\ part
II) it will become clear that the transport equations with
correlations already include the renormalization of one-particle
propagators in the dynamics. There is no need to calculate retarded
and advanced Green's functions independently of the distribution
functions.

As the dynamics of all quantities is just given by the Hamiltonian in
the form of an equation of motion (EOM) (Schr\"odinger or von-Neumann
equation), it seems to be more natural to consider only single-time
quantities. The problem to be solved is how to cut off the infinite
hierarchy of equations of motion in a consistent way. It will be shown
that the expansion of the expectation values of many-body operators in
terms of correlations (or, diagrammatically speaking, ``connected
parts'') allows a systematic decoupling of the hierarchy at any given
order. The lowest order (almost) agrees with the usual decoupling
procedure to obtain the Born approximation.

%
%
%
%
%

\section{Equations of motion and correlations}
\label{sec:EOM}

In this section the method will be explained for a fermionic system
with two-particle interaction in time-dependent external fields
described by the Hamiltonian 
\begin{equation}
\label{eq:Ham}
H = \sum_i \epsilon_i \psi_i^{\dag} \psi_i +
\sum_{i,j} h_{ij}^{\rm ext}(t) \psi_i^{\dag} \psi_j +
\frac{1}{2!} \cdot
\frac{1}{2!} \sum_{i_1,i_2 \atop j_1,j_2} v_{i_1 i_2, j_1 j_2}
\psi_{i_1}^{\dag} \psi_{i_2}^{\dag} \psi_{j_2} \psi_{j_1} = H_0^t + V
\quad.
\end{equation}
The fermionic annihilation and creation operators are denoted by
$\psi_i$ and $\psi_i^{\dag}$, the index $i$ referring to the
one-particle states. In the interaction term $V$ the matrix elements
$v_{i_1 i_2, j_1 j_2}$ are supposed to be anti-symmetric in the first
and second pair of indices ($ v_{i_2 i_1, j_1 j_2} = - v_{i_1 i_2, j_1
  j_2} = v_{i_1 i_2, j_2 j_1} $). The external fields are described by
the time-dependent hermitian matrix $h_{ij}^{\rm ext}(t)$. Sometimes
we use the notation $h_{ij}(t) =\delta_{ij} \epsilon_i +h_{ij}^{\rm
  ext}(t)$ for the one-particle Hamiltonian matrix.

The method is not restricted to this type of systems. It may as well
be applied to systems with bosons, even with ``anomalous'' (i.e.\ not
conserving the particle number) expectation values like $\langle b b
\rangle$. The necessary small alterations are mentioned from time to
time in this presentation and in part II the method is applied to an
electron-phonon system.

Throughout this work we use the Heisenberg representation, so that the
following equation of motion for operators holds:
\begin{equation}
  \label{eq:EOM}
  \frac{d}{dt} A(t) + i [A,H_0^t] (t) = - i [A,V] (t) \quad.
\end{equation}
For the case of an operator $A= \psi_{k_1} \cdots \psi_{k_n}
\psi_{k'_n}^{\dag} \cdots \psi_{k'_1}^{\dag}$ and the Hamiltonian of
Eq.~(\ref{eq:Ham}) the equation of motion reads
\begin{eqnarray}
  \lefteqn{ \frac{d}{dt} \left(\psi_{k_1} \cdots \psi_{k_n}
    \psi_{k'_n}^{\dag} \cdots 
    \psi_{k'_1}^{\dag}\right)(t) }
  \\
  &+& i \sum_j \left\{ h_{k_1 j}(t)  \left(\psi_{j}\psi_{k_2} \cdots
  \psi_{k_n} \psi_{k'_n}^{\dag} \cdots 
  \psi_{k'_1}^{\dag}\right)(t) + \cdots \right\}
  \nonumber \\
  &-& i \sum_i  \left\{ h_{i k'_1}(t)  \left(\psi_{k_1} \cdots \psi_{k_n}
  \psi_{k'_n}^{\dag} \cdots \psi_{k'_2}^{\dag}
  \psi_{i}^{\dag}\right)(t) + \cdots \right\}
  \nonumber \\ \nonumber 
  &=& -i \frac{1}{4}
  \sum_{i_1,i_2 \atop j_1,j_2} v_{i_1 i_2, j_1 j_2} \left[ \psi_{k_1}
  \cdots \psi_{k_n} \psi_{k'_n}^{\dag} \cdots \psi_{k'_1}^{\dag},
  \psi_{i_1}^{\dag} \psi_{i_2}^{\dag} \psi_{j_2} \psi_{j_1} \right] (t)
  \quad,
\end{eqnarray}
or more specifically for the case without external fields
\begin{eqnarray}
  \lefteqn{
    \left[ \frac{d}{dt} +i \left( \epsilon_{k_1} +\cdots +
      \epsilon_{k_n} - \epsilon_{k'_1} -\cdots -\epsilon_{k'_n}\right)
    \right] \left(\psi_{k_1} \cdots \psi_{k_n} \psi_{k'_n}^{\dag} \cdots
    \psi_{k'_1}^{\dag}\right)(t) }
  \nonumber \\
  &=& -i \frac{1}{4}
  \sum_{i_1,i_2 \atop j_1,j_2} v_{i_1 i_2, j_1 j_2} \left[ \psi_{k_1}
  \cdots \psi_{k_n} \psi_{k'_n}^{\dag} \cdots \psi_{k'_1}^{\dag},
  \psi_{i_1}^{\dag} \psi_{i_2}^{\dag} \psi_{j_2} \psi_{j_1} \right] (t)
  \quad.
\end{eqnarray}
The same equations of motion hold for the expectation values of these
multi-particle operators taken with an initial statistical operator
$\rho_0$ at time $t_0$, $\langle A \rangle_t = {\mathrm Tr} \left[
\rho_0 A(t) \right]$. (The Heisenberg representation of an operator is
supposed to be such that $A(t_0)=A$.) For our method to work $\rho_0$
does not have to be of any special form.  As the case of a
non-diagonal one-particle matrix $h_{ij}(t)$ is notationally
cumbersome, but does not pose any additional problems in the
derivation of the kinetic equations, we will carry out the derivation
for the case of a diagonal one-particle Hamiltonian, adding the
necessary alterations for non-vanishing external fields in the end.

Thus, given the initial state $\rho_0$ (mixed or pure) at time $t_0$,
we have to solve an infinite hierarchy of ordinary differential
equations (ODE). Of course, this is usually not feasible and some kind
of decoupling procedure is needed. Starting with one-particle
expectation values $\langle \psi_k^{\dag}\psi_{k'} \rangle_t$ the EOM
involve expectation values of two-particle operators. In the equation
of these quantities the ``collision term'' (the right hand side (rhs)
of Eq.~(\ref{eq:EOM})) is often approximated by factorizing all
expectation values into products of one-particle expectation values as
if the Wick theorem holds for an arbitrary state of the system. In
this way a non-Markovian Boltzmann equation in Born approximation is
obtained.

In this work we present a method to extend this procedure beyond the
Born approximation. It yields again an infinite hierarchy of EOM,
which is still exact, but now in terms of correlations. In contrast to
the usual hierarchy of EOM mentioned above, this hierarchy allows the
decoupling at any given order without any ambiguities.  The essential
ingredient is the following cluster expansion \cite{Sch} of products
of field operators, i.e.\ creation and annihilation operators, denoted
by $B_i$:
\begin{eqnarray}
  \label{eq:Cluster}
  \langle B_1 \rangle_t &=& \langle B_1 \rangle_t^c
  \quad, \nonumber
  \\
  \langle B_1 B_2 \rangle_t &=& \langle B_1 B_2 \rangle_t^c +
  \langle B_1 \rangle_t^c \cdot \langle B_2 \rangle_t^c 
  \quad, \\ 
  \langle B_1 B_2 B_3 \rangle_t &=& \langle B_1 B_2 B_3 \rangle_t^c +
  \langle B_1 B_2 \rangle_t^c \langle B_3 \rangle_t^c + \langle B_2
  \rangle_t^c \langle B_1 B_3 \rangle_t^c + \langle B_1 \rangle_t^c
  \langle B_2 B_3 \rangle_t^c + \langle B_1 \rangle_t^c \langle B_2
  \rangle_t^c \langle B_3 \rangle_t^c
  \nonumber \\ \vdots \nonumber{}
\end{eqnarray}
The $n$-th order correlations $\langle B_1 B_2 \cdots B_n \rangle_t^c$
are defined recursively.  The operators need not be normal-ordered.
Every expectation value $\langle B_1 B_2 \cdots B_n \rangle_t$ is
decomposed into sums over products of correlations in the following
way: The sum extends over all disjoint partitions of the set
$\{1,\dots,n\}$. The product is taken over the correlations
corresponding to each subset in the partition. Inside each correlation
the order of the field operators is retained. The overall sign of the
product is determined by the sign of the permutation of fermionic
field operators necessary to disentangle overlapping correlations. As
we assume that the initial statistical operator as well as the
Hamiltonian conserve the total number of fermions, only correlations
with an equal number of fermionic creation and annihilation operators
occur. Because of this property the sign of each decomposition into
correlations is well defined.

So far, the cluster expansion has been written down for both,
fermionic and bosonic operators. For fermionic systems the particle
number must be conserved and the usual rules for permutations of
fermionic operators apply. The one-particle correlations coincide with
the corresponding distribution functions
\begin{eqnarray}
  \label{1part_corr}
  \langle \psi_i \psi_j^{\dag} \rangle_t &=&
  \langle \psi_i \psi_j^{\dag} \rangle_t^c
  \quad,
  \\
  \langle \psi_j^{\dag} \psi_i \rangle_t &=&
  \langle \psi_j^{\dag} \psi_i \rangle_t^c
  \quad,
\end{eqnarray}
and the two-particle correlations $\langle \psi_{i_1} \psi_{i_2}
\psi_{j_2}^{\dag} \psi_{j_1}^{\dag} \rangle_t^c$, for example, are
defined by
\begin{equation}
  \label{2part_corr}
  \langle \psi_{i_1} \psi_{i_2} \psi_{j_2}^{\dag} \psi_{j_1}^{\dag}
  \rangle_t = 
  \langle \psi_{i_1} \psi_{i_2}
  \psi_{j_2}^{\dag} \psi_{j_1}^{\dag} \rangle_t^c
  + \langle \psi_{i_1}\psi_{j_1}^{\dag} \rangle_t^c
  \langle \psi_{i_2} \psi_{j_2}^{\dag} \rangle_t^c
  - \langle \psi_{i_1}\psi_{j_2}^{\dag} \rangle_t^c
  \langle \psi_{i_2} \psi_{j_1}^{\dag} \rangle_t^c
  \quad.
\end{equation}
From the canonical anti-commutation rules it follows immediately that
\begin{equation}
  \label{eq:commute}
  \langle \psi_i \psi_j^{\dag} \rangle_t^c = \delta_{i j} - \langle
  \psi_j^{\dag} \psi_i \rangle_t^c 
  \quad.
\end{equation}
In all higher-order correlations the fermionic operators anti-commute,
i.e.
\begin{eqnarray}
  \label{eq:commute_trivial}
  \langle \dots \psi_i \psi_j^{\dag} \dots \rangle_t^c = - \langle
  \dots \psi_j^{\dag} \psi_i \dots \rangle_t^c 
  \quad, \\ \nonumber{}
  \langle \dots \psi_i \psi_j \dots \rangle_t^c = - \langle \dots
  \psi_j \psi_i \dots \rangle_t^c 
  \quad,
\end{eqnarray}
which can be easily proved by induction.  Therefore, it is sufficient
to consider the following anti-normal ordered correlations:
\begin{equation}
  \label{eq:corr_fcn}
  \langle \psi_{k_1} \cdots \psi_{k_n} \psi_{k'_n}^{\dag} \cdots
  \psi_{k'_1}^{\dag} \rangle_t^c \quad,\ n \geq 1\quad.
\end{equation}
We will call the one-particle correlations ($n=1$) contractions.

Any expectation value of the form $ \langle \psi_{k_1} \cdots
\psi_{k_n} \psi_{k'_n}^{\dag} \cdots \psi_{k'_1}^{\dag} \rangle_t$ can
be expressed in terms of correlations as can be seen directly from the
definition of the correlations in Eq.~(\ref{eq:Cluster}). This enables
us to transform the infinite hierarchy of EOM into an equivalent one
in terms of correlations. So far, nothing seems to be gained. But we
will see that the EOM for the correlations allow an easy and
unambiguous breaking of the hierarchy at any desired order, in
contrast to the usual EOM. The lowest order is usually of the type of
a non-Markovian Boltzmann equation in Born approximation.

In the next section we derive diagrams for the correlations which are
as convenient as usual Feynman graphs for organizing the calculations.
It turns out that the temporal evolution of correlations is determined
by ``connected'' diagrams only.

We start with the EOM for the one-particle distribution function
\begin{eqnarray}
  \label{eq:one-particle}
  \lefteqn{
    \left[ \frac{d}{dt} +i(\epsilon_{k_1} -\epsilon_{k'_1})
    \right] \langle \psi_{k_1} \psi_{k'_1}^{\dag} \rangle_t^c } &&
  \\ \nonumber{}
  &=& -i \frac{1}{4} \sum_{i_1,i_2 \atop j_1,j_2} v_{i_1 i_2, j_1 j_2}
  \left\{ \langle \psi_{k_1} \psi_{k'_1}^{\dag} \psi_{i_1}^{\dag}
    \psi_{i_2}^{\dag} \psi_{j_2} \psi_{j_1} \rangle_t - \langle
    \psi_{i_1}^{\dag} \psi_{i_2}^{\dag} \psi_{j_2} \psi_{j_1}
    \psi_{k_1} \psi_{k'_1}^{\dag} \rangle_t
  \right\}
  \\  \nonumber{}
  &=& -i \sum_{i_1,i_2 \atop j_1,j_2} v_{i_1 i_2, j_1 j_2} 
  \left\{
    -\frac{1}{2} \left(\langle \psi_{k_1} \psi_{i_1}^{\dag} \rangle_t^c
    + \langle\psi_{i_1}^{\dag}\psi_{k_1}\rangle_t^c\right) \cdot \langle
    \psi_{j_1}\psi_{j_2}\psi_{i_2}^{\dag}\psi_{k'_1}^{\dag}\rangle_t^c
  \right.
  \\ \nonumber{}
  && +\frac{1}{2}
  \left(\langle\psi_{k'_1}^{\dag}\psi_{j_1}\rangle_t^c
    +\langle\psi_{j_1}\psi_{k'_1}^{\dag}\rangle_t^c
  \right) \cdot
  \langle
  \psi_{k_1}\psi_{j_2}\psi_{i_2}^{\dag}\psi_{i_1}^{\dag}\rangle_t^c
  \\ && \nonumber{}
  \left.  +\left(-\langle\psi_{k_1}\psi_{i_1}^{\dag}\rangle_t^c
    \langle\psi_{k'_1}^{\dag}\psi_{j_1}\rangle_t^c +
    \langle\psi_{i_1}^{\dag}\psi_{k_1}\rangle_t^c
    \langle\psi_{j_1}\psi_{k'_1}^{\dag}\rangle_t^c\right) \cdot
    \langle\psi_{i_2}^{\dag}\psi_{j_2}\rangle_t^c
  \right\}
\quad.
\end{eqnarray}
The reader might wonder why in the first step we did not calculate
explicitly the commutator leaving only terms with four instead of six
fermionic operators. It turns out that the cluster expansion and the
derivation of diagrammatic rules is easier when one leaves the
commutator as it is. Of course, after the cluster expansion of the
commutator the correlations with six field operators cancel because of
the rule Eq.~(\ref{eq:commute_trivial}).  Only terms with at least one
contraction remain because the non-trivial part of the fermionic
anti-commutation rules shows up only in the contractions. In addition,
we simplified the result by combining terms differing only in the
exchange of the two annihilation or creation operators of the
interaction term. This is possible as we use anti-symmetrized
interaction matrix elements. In this way the prefactor of $1/4 =
1/{2!} \cdot 1/{2!}$ is sometimes replaced by $1/{2!}$ or 1. Later, we
will give a simple diagrammatic rule to determine this so-called
symmetry factor.  For pedagogical reasons we have not simplified
$\langle\psi_{i}^{\dag}\psi_{j}\rangle_t^c
+\langle\psi_{j}\psi_{i}^{\dag}\rangle_t^c = \delta_{ij}$ as we will
do from now on.

In the next step, the EOM for the newly encountered correlations
$\langle \psi_{k_1} \psi_{k_2} \psi_{k'_2}^{\dag} \psi_{k'_1}^{\dag}
\rangle_t^c$ is needed. It can be calculated in the following way: By
definition
\begin{equation}
  \label{eq:2part-corr}
  \langle \psi_{k_1} \psi_{k_2} \psi_{k'_2}^{\dag} \psi_{k'_1}^{\dag}
  \rangle_t^c = \langle \psi_{k_1} \psi_{k_2} \psi_{k'_2}^{\dag}
  \psi_{k'_1}^{\dag} \rangle_t - \langle \psi_{k_1}
  \psi_{k'_1}^{\dag}\rangle^c_t \langle \psi_{k_2} \psi_{k'_2}^{\dag}
  \rangle^c_t + \langle \psi_{k_1} \psi_{k'_2}^{\dag}\rangle^c_t \langle
  \psi_{k_2} \psi_{k'_1}^{\dag} \rangle^c_t \quad,
\end{equation}
so that the EOM for $ \langle \psi_{k_1} \psi_{k_2} \psi_{k'_2}^{\dag}
\psi_{k'_1}^{\dag} \rangle_t$ is needed. It reads
\begin{eqnarray}
  \label{eq:2partEOM}
  \lefteqn{ \left[ \frac{d}{dt} +i (\epsilon_{k_1} + \epsilon_{k_2} -
    \epsilon_{k'_1} - \epsilon_{k'_2} ) \right] \langle
    \psi_{k_1}\psi_{k_2}\psi_{k'_2}^{\dag} \psi_{k'_1}^{\dag}
    \rangle_t }
  \\ \nonumber
  &=& -i \sum_{i_1,i_2 \atop j_1,j_2}
  v_{i_1 i_2, j_1 j_2} \left\{ -\frac{1}{2} \delta_{i_1,k_1} \langle
  \psi_{k_2}\psi_{j_2} \psi_{j_1} \psi_{k'_1}^{\dag} \psi_{i_2}^{\dag}
  \psi_{k'_2}^{\dag} \rangle_t^c + (k_1 \leftrightarrow k_2) \right.
  \\ \nonumber &&
  + \frac{1}{2} \delta_{j_1,k'_1} \langle \psi_{k_2}\psi_{j_2}
  \psi_{k_1} \psi_{i_1}^{\dag} \psi_{i_2}^{\dag} \psi_{k'_2}^{\dag}
  \rangle_t^c - (k'_1 \leftrightarrow k'_2)
  \\ \nonumber &&
  + \frac{1}{2} \left(\langle \psi_{k_1}\psi_{i_1}^{\dag}
  \rangle_t^c\langle \psi_{k_2}\psi_{i_2}^{\dag} \rangle_t^c - \langle
  \psi_{i_1}^{\dag} \psi_{k_1} \rangle_t^c\langle \psi_{i_2}^{\dag}
  \psi_{k_2} \rangle_t^c\right) \langle
  \psi_{j_1}\psi_{j_2}\psi_{k'_2}^{\dag} \psi_{k'_1}^{\dag}
  \rangle_t^c
  \\ \nonumber &&
  + \frac{1}{2} \left(\langle \psi_{k'_1}^{\dag} \psi_{j_1}
  \rangle_t^c\langle \psi_{k'_2}^{\dag} \psi_{j_2} \rangle_t^c -
  \langle \psi_{j_1}\psi_{k'_1}^{\dag} \rangle_t^c\langle
  \psi_{j_2}\psi_{k'_2}^{\dag} \rangle_t^c\right) \langle
  \psi_{k_1}\psi_{k_2}\psi_{i_2}^{\dag} \psi_{i_1}^{\dag} \rangle_t^c
  \\ \nonumber &&
  + 
  \left[ \left(\langle \psi_{k_1}\psi_{i_1}^{\dag}
    \rangle_t^c\langle \psi_{k'_1}^{\dag} \psi_{j_1} \rangle_t^c -
    \langle \psi_{i_1}^{\dag} \psi_{k_1} \rangle_t^c\langle
    \psi_{j_1}\psi_{k'_1}^{\dag} \rangle_t^c\right) \langle
    \psi_{k_2}\psi_{j_2}\psi_{i_2}^{\dag} \psi_{k'_2}^{\dag} \rangle_t^c
  \right.
  \\ \nonumber &&
  - \frac{1}{2} \delta_{i_1,k_1} \langle \psi_{k_2}\psi_{k'_2}^{\dag}
  \rangle_t^c \langle \psi_{j_1} \psi_{j_2}\psi_{i_2}^{\dag}
  \psi_{k'_1}^{\dag} \rangle_t^c + \frac{1}{2} \delta_{j_1,k'_1}
  \langle \psi_{k_2}\psi_{k'_2}^{\dag} \rangle_t^c \langle \psi_{k_1}
  \psi_{j_2}\psi_{i_2}^{\dag} \psi_{i_1}^{\dag} \rangle_t^c
  \\ \nonumber &&
  \left.  + \langle \psi_{k_2}\psi_{k'_2}^{\dag} \rangle_t^c
    \left(-\langle \psi_{k_1}\psi_{i_1}^{\dag} \rangle_t^c \langle
      \psi_{k'_1}^{\dag} \psi_{j_1} \rangle_t^c + \langle
      \psi_{i_1}^{\dag} \psi_{k_1} \rangle_t^c \langle
      \psi_{j_1}\psi_{k'_1}^{\dag} \rangle_t^c
    \right)
    \langle \psi_{i_2}^{\dag} \psi_{j_2} \rangle_t^c 
  \right]
  \\ \nonumber &&
  - (k_1 \leftrightarrow k_2) - (k'_1 \leftrightarrow k'_2) + (k_1
  \leftrightarrow k_2 , k'_1 \leftrightarrow k'_2)
  \\ \nonumber &&
  + \delta_{i_1,k_1} \langle \psi_{i_2}^{\dag} \psi_{j_2} \rangle_t^c
  \langle \psi_{j_1} \psi_{k_2}\psi_{k'_2}^{\dag} \psi_{k'_1}^{\dag}
  \rangle_t^c - (k_1 \leftrightarrow k_2)
  \\ \nonumber &&
  - \delta_{j_1,k'_1} \langle \psi_{i_2}^{\dag} \psi_{j_2} \rangle_t^c
  \langle \psi_{k_1} \psi_{k_2}\psi_{k'_2}^{\dag} \psi_{i_1}^{\dag}
  \rangle_t^c + (k'_1 \leftrightarrow k'_2)
  \\ \nonumber &&
  + \left( \langle \psi_{k_1}\psi_{i_1}^{\dag} \rangle_t^c\langle
  \psi_{k_2}\psi_{i_2}^{\dag} \rangle_t^c \langle \psi_{k'_1}^{\dag}
  \psi_{j_1} \rangle_t^c\langle \psi_{k'_2}^{\dag} \psi_{j_2}
  \rangle_t^c \right.
  \\ \nonumber{} &&
  \left.  \left.  - \langle \psi_{i_1}^{\dag} \psi_{k_1}
    \rangle_t^c\langle \psi_{i_2}^{\dag} \psi_{k_2} \rangle_t^c
    \langle \psi_{j_1}\psi_{k'_1}^{\dag} \rangle_t^c\langle
    \psi_{j_2}\psi_{k'_2}^{\dag} \rangle_t^c \right) 
  \right\} \quad,
\end{eqnarray}
where we have already used the cluster expansion for the ``collision
term'', i.e.\ for the rhs of the equation. From this equation,
Eq.~(\ref{eq:one-particle}), and the definition of the corresponding
correlation Eq.~(\ref{eq:2part-corr}), we obtain the following EOM for
$\langle \psi_{k_1} \psi_{k_2} \psi_{k'_2}^{\dag} \psi_{k'_1}^{\dag}
\rangle_t^c$:
\begin{eqnarray}
  \label{eq:2part-corrEOM}
  \lefteqn{ \left[ \frac{d}{dt} +i (\epsilon_{k_1} + \epsilon_{k_2} -
    \epsilon_{k'_1} - \epsilon_{k'_2} ) \right] \langle
    \psi_{k_1}\psi_{k_2}\psi_{k'_2}^{\dag} \psi_{k'_1}^{\dag}
    \rangle_t^c }
  \\ \nonumber
  &=& \left[ \frac{d}{dt} +i
  (\epsilon_{k_1} + \epsilon_{k_2} - \epsilon_{k'_1} - \epsilon_{k'_2}
  ) \right] \langle \psi_{k_1}\psi_{k_2}\psi_{k'_2}^{\dag}
  \psi_{k'_1}^{\dag} \rangle_t
  \\ \nonumber &&
  - \left\{ \langle \psi_{k_1}\psi_{k'_1}^{\dag} \rangle_t \left[
  \frac{d}{dt} +i(\epsilon_{k_2}-\epsilon_{k'_2}) \right] \langle
  \psi_{k_2}\psi_{k'_2}^{\dag} \rangle_t \right.
  \\ \nonumber{} &&
  \left.  + \langle \psi_{k_2}\psi_{k'_2}^{\dag} \rangle_t \left[
    \frac{d}{dt} +i(\epsilon_{k_1}-\epsilon_{k'_1}) \right] \langle
    \psi_{k_1}\psi_{k'_1}^{\dag} \rangle_t
  \right\}
  \\ \nonumber&&
  + \left\{ \langle \psi_{k_1}\psi_{k'_2}^{\dag} \rangle_t \left[
  \frac{d}{dt} +i(\epsilon_{k_2}-\epsilon_{k'_1}) \right] \langle
  \psi_{k_2}\psi_{k'_1}^{\dag} \rangle_t \right.
  \\ \nonumber{} &&
  \left.  + \langle \psi_{k_2}\psi_{k'_1}^{\dag} \rangle_t \left[
    \frac{d}{dt} +i(\epsilon_{k_1}-\epsilon_{k'_2}) \right] \langle
    \psi_{k_1}\psi_{k'_2}^{\dag} \rangle_t 
  \right\}
  \\ \nonumber 
  &=& -i \sum_{i_1,i_2 \atop j_1,j_2} v_{i_1 i_2, j_1 j_2} \left\{
  -\frac{1}{2} \delta_{i_1,k_1} \langle \psi_{k_2}\psi_{j_2}\psi_{j_1}
  \psi_{k'_1}^{\dag} \psi_{i_2}^{\dag} \psi_{k'_2}^{\dag} \rangle_t^c
  +( k_1 \leftrightarrow k_2) \right.
  \\ \nonumber &&
  + \frac{1}{2} \delta_{j_1,k'_1} \langle
  \psi_{k_2}\psi_{j_2}\psi_{k_1} \psi_{i_1}^{\dag} \psi_{i_2}^{\dag}
  \psi_{k'_2}^{\dag} \rangle_t^c -( k'_1 \leftrightarrow k'_2)
  \\ \nonumber &&
  + \frac{1}{2} \left(\langle \psi_{k_1}\psi_{i_1}^{\dag} \rangle_t^c
  \langle \psi_{k_2}\psi_{i_2}^{\dag} \rangle_t^c - \langle
  \psi_{i_1}^{\dag} \psi_{k_1}\rangle_t^c \langle \psi_{i_2}^{\dag}
  \psi_{k_2}\rangle_t^c \right) \langle
  \psi_{j_1}\psi_{j_2}\psi_{k'_2}^{\dag} \psi_{k'_1}^{\dag}
  \rangle_t^c
  \\ \nonumber &&
  + \frac{1}{2} \left(\langle \psi_{k'_1}^{\dag} \psi_{j_1}\rangle_t^c
  \langle \psi_{k'_2}^{\dag} \psi_{j_2}\rangle_t^c - \langle
  \psi_{j_1}\psi_{k'_1}^{\dag} \rangle_t^c \langle
  \psi_{j_2}\psi_{k'_2}^{\dag} \rangle_t^c \right) \langle
  \psi_{k_1}\psi_{k_2}\psi_{i_2}^{\dag} \psi_{i_1}^{\dag} \rangle_t^c
  \\ \nonumber &&
  + \left(\langle \psi_{k_1}\psi_{i_1}^{\dag} \rangle_t^c \langle
  \psi_{k'_1}^{\dag} \psi_{j_1}\rangle_t^c - \langle \psi_{i_1}^{\dag}
  \psi_{k_1}\rangle_t^c \langle \psi_{j_1}\psi_{k'_1}^{\dag}
  \rangle_t^c \right) \langle \psi_{k_2}\psi_{j_2}\psi_{i_2}^{\dag}
  \psi_{k'_2}^{\dag} \rangle_t^c
  \\ \nonumber &&
  - (k_1 \leftrightarrow k_2) - (k'_1 \leftrightarrow k'_2) + (k_1
  \leftrightarrow k_2 , k'_1 \leftrightarrow k'_2)
  \\ \nonumber &&
  + \delta_{i_1,k_1} \langle \psi_{i_2}^{\dag} \psi_{j_2}\rangle_t^c
  \langle \psi_{j_1} \psi_{k_2}\psi_{k'_2}^{\dag} \psi_{k'_1}^{\dag}
  \rangle_t^c - (k_1 \leftrightarrow k_2)
  \\ \nonumber &&
  - \delta_{j_1,k'_1} \langle \psi_{i_2}^{\dag} \psi_{j_2}\rangle_t^c
  \langle \psi_{k_1} \psi_{k_2}\psi_{k'_2}^{\dag} \psi_{i_1}^{\dag}
  \rangle_t^c + (k'_1 \leftrightarrow k'_2)
  \\ \nonumber &&
  + \left( \langle \psi_{k_1}\psi_{i_1}^{\dag} \rangle_t^c \langle
  \psi_{k_2}\psi_{i_2}^{\dag} \rangle_t^c \langle \psi_{k'_1}^{\dag}
  \psi_{j_1}\rangle_t^c \langle \psi_{k'_2}^{\dag}
  \psi_{j_2}\rangle_t^c \right.
  \\ \nonumber{} &&
  \left.  
    \left.  - \langle \psi_{i_1}^{\dag} \psi_{k_1}\rangle_t^c
      \langle \psi_{i_2}^{\dag} \psi_{k_2}\rangle_t^c \langle
      \psi_{j_1}\psi_{k'_1}^{\dag} \rangle_t^c \langle
      \psi_{j_2}\psi_{k'_2}^{\dag} \rangle_t^c 
    \right) 
  \right\} \quad.
\end{eqnarray}
We would like to mention that the diagrams presented in the next
section facilitate the determination of the EOM in quite the same way
as Feynman graphs do in equilibrium theory. Note that the EOM for the
correlation is the same as the EOM for the corresponding $n$-point
function with just some terms missing. Diagrammatically these terms
correspond to unconnected graphs.

Now we want to describe the procedure for the breaking of the
hierarchy of EOM. With the two-particle interaction the EOM for a
given correlation of order $n$ contains correlations of order $l \leq
n+1$. If a calculation up to order $n$ is desired, we just neglect the
correlations of order $n+1$ in the EOM for the correlations of order
$n$ and get a closed system of ODE for the correlations of order $l
\leq n$. This procedure has the advantage of working without
ambiguities at every given order. In addition, the transport equations
are already of the form in which they are usually solved, i.e.\ as a
system of ODE. This system allows for the inclusion of initial
correlations up to order $n$, that means that additional information
about the initial state $\rho_0$ can be included compared with the
usual approaches. It should be clear that our approximation method is
in some sense an expansion in powers of $V\cdot \Delta t$, where $V$
is the typical interaction strength and $\Delta t = t - t_0$ is the
time passed since the initial time. But at the given order it is the
best possible approximation for the one-particle distribution
function. No resummation as in ordinary perturbation theory \cite{Sch}
is needed.

The same procedure does not work for the usual hierarchy of EOM
because the EOM for the $2n$-point function $\langle \psi_{k_1} \cdots
\psi_{k_n} \psi_{k'_n}^{\dag} \cdots \psi_{k'_1}^{\dag} \rangle_t$
essentially involves only $2(n+1)$-point functions. Therefore, these
functions cannot just be neglected. Some kind of decoupling mechanism
is needed.  This was done e.g.\ by Zimmermann and Wauer \cite{Zi} in
one order higher than Born approximation for a Jaynes-Cummings model.
But as the only consistent way of doing it is in terms of a cluster
expansion, it is best to start with correlations from the beginning.

Another noteworthy feature of our method is the fact that the
derivatives $\left( \frac{d}{dt} \right)^l \langle \psi_{k_1}
\psi_{k'_1}^{\dag} \rangle_t$ at the initial time $t=t_0$ are
correctly described for $0 \leq l \leq n-1$ when the cut-off is at
order $n$. A further advantage of using correlations is the fact that
they vanish for all orders $n \geq 2$ in the case of a non-interacting
system in a grand-canonical ensemble (Wick's theorem \cite{Wick}), in
contrast to $n$-point functions. For this reason the EOM for
correlations yield better results for systems with weak interaction
showing relaxation into thermal equilibrium (cf.\ part II). The most
convincing argument in favor of this method are the very promising
results in comparison with the exact model calculations presented in
II for one-dimensional electron-phonon systems. In three dimensions
the order of the system of ODE is already huge at the level of the
Born approximation so that it seems not feasible to go even one order
higher.  Even in a homogeneous system with momentum conservation, for
example, the $n$-th order correlation has $2n-1$ free momentum
indices, i.e.\ the number of ODE is of the order $N^{(2n-1)d}$, where
$N^d$ is the number of one-electron states. Because of this fact
numerical calculations in three dimensions are often performed
assuming an isotropic distribution in $\vec{k}$-space, thus reducing
the effective dimension. With electron-phonon instead of
electron-electron interaction, the situation is slightly better,
because the usual electron-phonon interaction term contains only
products of three instead of four field operators.

Before turning to diagrams, we discuss the kinetic equations in the
lowest approximations. Neglecting all correlations of order $n \geq 2$
we obtain
\begin{equation}
  \label{eq:HF}
  \frac{d}{dt} \langle \psi_{k_1}\psi_{k'_1}^{\dag} \rangle_t^c +i
  \sum_j \widetilde{\epsilon}_{k_1 j}(t) \langle
  \psi_{j}\psi_{k'_1}^{\dag} \rangle_t^c - i \sum_i \langle
  \psi_{k_1}\psi_{i}^{\dag} \rangle_t^c\, \widetilde{\epsilon}_{i
    k'_1}(t) = 0 \quad,
\end{equation}
which is just the propagation in the external fields with the
one-particle energies corrected by the time-dependent Hartree-Fock
energies $v_{ij}^{\rm HF}(t) := \sum_{i_2,j_2} v_{i i_2, j j_2}
\langle \psi_{i_2}^{\dag} \psi_{j_2} \rangle_t^c$,
$\widetilde{\epsilon}_{ij}(t) := \delta_{ij} \epsilon_{i} +
h_{ij}^{\rm ext}(t) + v_{ij}^{\rm HF}(t)$.  Note that we have added
the contribution of the external fields again.  This equation does not
contain a ``collision term''. At least the next order has to be
retained. We obtain the following closed system of ODE:
\begin{eqnarray}
  \label{eq:Order2}
  \lefteqn{ \frac{d}{dt} \langle
    \psi_{k_1}\psi_{k'_1}^{\dag}\rangle_t^c +i \sum_j
    \widetilde{\epsilon}_{k_1 j}(t) \langle \psi_{j}
    \psi_{k'_1}^{\dag} \rangle_t^c - i \sum_i \langle \psi_{k_1}
    \psi_{i}^{\dag} \rangle_t^c\, \widetilde{\epsilon}_{i k'_1}(t) }
  \\ \nonumber{}
  &=& \frac{i}{2} \left\{ \sum_{i_2, j_1, j_2} v_{k_1 i_2, j_1 j_2}
  \langle \psi_{j_1} \psi_{j_2} \psi_{i_2}^{\dag} \psi_{k'_1}^{\dag}
  \rangle_t^c -\sum_{i_1, i_2, j_2} v_{i_1 i_2, k'_1 j_2} \langle
  \psi_{k_1} \psi_{j_2} \psi_{i_2}^{\dag} \psi_{i_1}^{\dag}
  \rangle_t^c \right\} \quad,
\end{eqnarray}
\begin{eqnarray}
  \label{eq:Order_2}
  \frac{d}{dt} \langle \psi_{k_1} \psi_{k_2}
  \psi_{k'_2}^{\dag} \psi_{k'_1}^{\dag} \rangle_t^c 
  &+& i\sum_j
  \left\{ \widetilde{\epsilon}_{k_1 j}(t) \langle \psi_{j}
    \psi_{k_2} \psi_{k'_2}^{\dag} \psi_{k'_1}^{\dag} \rangle_t^c +
    \widetilde{\epsilon}_{k_2 j}(t) \langle \psi_{k_1} \psi_{j}
    \psi_{k'_2}^{\dag} \psi_{k'_1}^{\dag} \rangle_t^c
  \right\}
  \\ \nonumber{}
  &-& i \sum_i \left\{ \langle \psi_{k_1}
  \psi_{k_2} \psi_{k'_2}^{\dag} \psi_{i}^{\dag} \rangle_t^c \,
  \widetilde{\epsilon}_{i k'_1}(t) + \langle \psi_{k_1} \psi_{k_2}
  \psi_{i}^{\dag} \psi_{k'_1}^{\dag} \rangle_t^c \,
  \widetilde{\epsilon}_{i k'_2}(t) \right\} 
  \\ \nonumber{}
  &+& \frac{i}{2} \left( \sum_{j_1 j_2} V_{k_1 k_2, j_1
    j_2}(t) \langle \psi_{j_1} \psi_{j_2} \psi_{k'_2}^{\dag}
  \psi_{k'_1}^{\dag} \rangle_t^c -\sum_{i_1 i_2} (V_{k'_1 k'_2, i_1
    i_2}(t))^* \langle \psi_{k_1} \psi_{k_2} \psi_{i_2}^{\dag}
  \psi_{i_1}^{\dag} \rangle_t^c \right) 
  \\ \nonumber{}
  &+& i \sum_{i_2 j_2} W_{k_1 i_2, k'_1 j_2}(t) \langle
  \psi_{k_2} \psi_{j_2} \psi_{i_2}^{\dag} \psi_{k'_2}^{\dag}
  \rangle_t^c - (k_1 \leftrightarrow k_2) - (k'_1 \leftrightarrow
  k'_2)
  \\ \nonumber{}
  && {}+ (k_1 \leftrightarrow k_2 , k'_1 \leftrightarrow k'_2) 
  \\ \nonumber{}
  = &-& i \sum_{i_1,i_2 \atop j_1,j_2} v_{i_1 i_2, j_1 j_2} \left(
  \langle \psi_{k_1}\psi_{i_1}^{\dag}\rangle_t^c \langle
  \psi_{k_2}\psi_{i_2}^{\dag}\rangle_t^c \langle
  \psi_{k'_1}^{\dag}\psi_{j_1}\rangle_t^c \langle
  \psi_{k'_2}^{\dag}\psi_{j_2}\rangle_t^c \right.
  \\ \nonumber &&
  \left.  {}- \langle \psi_{i_1}^{\dag}\psi_{k_1}\rangle_t^c \langle
    \psi_{i_2}^{\dag}\psi_{k_2}\rangle_t^c \langle
    \psi_{j_1}\psi_{k'_1}^{\dag}\rangle_t^c \langle
    \psi_{j_2}\psi_{k'_2}^{\dag}\rangle_t^c 
  \right) \quad.
\end{eqnarray}
The time-dependent Hartree-Fock corrections of the one-particle
energies as well as the terms with
\begin{equation}
  \label{eq:V}
  V_{k_1 k_2, j_1 j_2}(t) = \sum_{i_1 i_2} v_{i_1 i_2, j_1 j_2} \left(
  \langle \psi_{k_1}\psi_{i_1}^{\dag}\rangle_t^c \langle
  \psi_{k_2}\psi_{i_2}^{\dag}\rangle_t^c -\langle
  \psi_{i_1}^{\dag}\psi_{k_1}\rangle_t^c \langle
  \psi_{i_2}^{\dag}\psi_{k_2}\rangle_t^c \right)
\end{equation}
and
\begin{equation}
  \label{eq:W}
  W_{k_1 i_2, k'_1 j_2}(t) = \sum_{i_1 j_1} v_{i_1 i_2, j_1 j_2}
  \left( \langle \psi_{k_1}\psi_{i_1}^{\dag}\rangle_t^c \langle
    \psi_{k'_1}^{\dag}\psi_{j_1}\rangle_t^c -\langle
    \psi_{i_1}^{\dag}\psi_{k_1}\rangle_t^c \langle
    \psi_{j_1}\psi_{k'_1}^{\dag}\rangle_t^c 
  \right) \quad,
\end{equation}
which correct the correlated propagation of two particles, are not
obtained by a naive decoupling procedure for the two-particle
expectation values.

For the special case of a homogeneous, spin-independent initial state
$\rho_0$ and a momentum and spin conserving Hamiltonian without
external fields, the EOM simplify considerably. We denote the
one-particle distribution function by $n_k(t) := \langle
\psi_{k}^{\dag} \psi_{k}\rangle_t^c $ with $k=(\vec{k},\sigma)$.
\begin{equation}
  \label{1part-hom}
  \frac{d}{dt} n_k(t) = -\frac{i}{2}
  \left\{ \sum_{k_2 k'_1
      k'_2} v_{k k_2, k'_1 k'_2} \langle \psi_{k'_1} \psi_{k'_2}
    \psi_{k_2}^{\dag} \psi_{k}^{\dag} \rangle_t^c - \mathrm{c.c.}
  \right\} 
  \quad,
\end{equation}
\begin{eqnarray} 
  \lefteqn{ \left[ \frac{d}{dt} + i (\widetilde{\epsilon}_{k_1}(t) +
    \widetilde{\epsilon}_{k_2}(t) - \widetilde{\epsilon}_{k'_1}(t) -
    \widetilde{\epsilon}_{k'_2}(t)) \right] \langle \psi_{k_1}
    \psi_{k_2} \psi_{k'_2}^{\dag} \psi_{k'_1}^{\dag} \rangle_t^c }
  \\ \nonumber{}
  && {} +\frac{i}{2} \sum_{j_1 j_2} v_{k_1 k_2, j_1 j_2} (1 -
    n_{k_1}(t) -n_{k_2}(t)) \langle \psi_{j_1} \psi_{j_2}
    \psi_{k'_2}^{\dag} \psi_{k'_1}^{\dag} \rangle_t^c 
  \\ \nonumber{}
  && {} -\frac{i}{2} \sum_{i_1 i_2} v_{i_1 i_2, k'_1 k'_2} (1 -
    n_{k'_1}(t) -n_{k'_2}(t)) \langle \psi_{k_1} \psi_{k_2}
    \psi_{i_2}^{\dag} \psi_{i_1}^{\dag} \rangle_t^c 
  \\ \nonumber{}
  && {} +i \sum_{i_2 j_2} v_{k_1 i_2, k'_1 j_2} (n_{k'_1}(t)
    -n_{k_1}(t)) \langle \psi_{k_2} \psi_{j_2} \psi_{i_2}^{\dag}
    \psi_{k'_2}^{\dag} \rangle_t^c 
  \\ \nonumber{}
  && {} - (k_1 \leftrightarrow k_2) - (k'_1 \leftrightarrow k'_2)
    + (k_1 \leftrightarrow k_2 , k'_1 \leftrightarrow k'_2) 
  \\ \nonumber{}
  &=& -i v_{k_1 k_2, k'_1 k'_2} \left( (1-n_{k_1}(t))(1-n_{k_2}(t))
  n_{k'_1}(t) n_{k'_2}(t) \right.
  \\ \nonumber{} &&
  \left. {} - n_{k_1}(t) n_{k_2}(t)(1-n_{k'_1}(t))(1-n_{k'_2}(t))
  \right) \quad.
\end{eqnarray}
In this form, the ODE are solved numerically.  Neglecting the
Hartree-Fock contribution and the other corrections on the lhs of the
second equation, this EOM can easily be integrated:
\begin{eqnarray}
  \langle \psi_{k_1} \psi_{k_2} \psi_{k'_2}^{\dag} \psi_{k'_1}^{\dag}
  \rangle_t^c &=& \langle \psi_{k_1}\psi_{k_2}\psi_{k'_2}^{\dag}
  \psi_{k'_1}^{\dag} \rangle_{t_0}^c \, e^{-i(\epsilon_{k_1}
    +\epsilon_{k_2} - \epsilon_{k'_1} - \epsilon_{k'_2})(t-t_0)}
  \\ \nonumber{} &&
  -i v_{k_1 k_2, k'_1 k'_2} \int_{t_0}^{t}dt'\, e^{-i(\epsilon_{k_1}
    +\epsilon_{k_2} - \epsilon_{k'_1} - \epsilon_{k'_2})(t-t')}
  \\ \nonumber{} &&
  \times \left( (1-n_{k_1})(1-n_{k_2}) n_{k'_1} n_{k'_2} - n_{k_1}
  n_{k_2}(1-n_{k'_1})(1-n_{k'_2}) \right)({t'}) \quad.
\end{eqnarray}
In the case of no initial correlations, the transport equation for the
one-particle distribution function can then be written as an
integro-differential equation:
\begin{eqnarray}
  \frac{d}{dt} n_k(t) &=& - \sum_{k_2 k'_1 k'_2} |v_{k k_2, k'_1
    k'_2}|^2 \int_{t_0}^{t}dt'\, \cos ((\epsilon_{k'_1} +\epsilon_{k'_2} -
  \epsilon_{k} - \epsilon_{k_2})(t-t'))
  \\ \nonumber{} &&
  \times \left(  n_{k} n_{k_2}(1-n_{k'_1})(1-n_{k'_2}) 
  - (1-n_{k})(1-n_{k_2}) n_{k'_1} n_{k'_2} \right)(t') \quad.
\end{eqnarray}

For a local two-particle interaction ($k_1 = (\vec{k_1},\sigma_1)$,
etc.)
\begin{equation}
  \label{local-inter}
  v_{k_1 k_2, k'_1 k'_2} = \left\{ \tilde{v}_{\vec{k_1}-\vec{k'_1}}
  \delta_{\sigma_1 \sigma'_1} \delta_{\sigma_2 \sigma'_2} -
  \tilde{v}_{\vec{k_1}-\vec{k'_2}} \delta_{\sigma_1 \sigma'_2}
  \delta_{\sigma_2 \sigma'_1} \right\}
  \delta_{\vec{k_1}+\vec{k_2},\vec{k'_1}+\vec{k'_2}}
\end{equation}
and for an initially spin-independent distribution ($n_{\vec{k}} (t)
:= n_{\vec{k},\sigma}(t)$) we obtain
\begin{eqnarray}
  \label{ee-Born}
  \frac{d}{dt} n_{\vec{k}}(t) &=& - 2\sum_{\vec{k_2} \vec{k'_1}
    \vec{k'_2}} \delta_{\vec{k}+\vec{k_2},\vec{k'_1}+\vec{k'_2}}
  \left\{ 2(\tilde{v}_{\vec{k}-\vec{k'_1}})^2 -
    \tilde{v}_{\vec{k}-\vec{k'_1}} \tilde{v}_{\vec{k}-\vec{k'_2}}
  \right\}
  \\ \nonumber{} &&
  \times \int_{t_0}^{t}dt'\, \cos ((\epsilon_{k'_1} +\epsilon_{k'_2} -
  \epsilon_{k} - \epsilon_{k_2})(t-t'))
  \\ \nonumber{} &&
  \times \left( n_{\vec{k}} n_{\vec{k_2}}(1-n_{\vec{k'_1}})(1-n_{\vec{k'_2}})
  - (1-n_{\vec{k}})(1-n_{\vec{k_2}}) n_{\vec{k'_1}}
  n_{\vec{k'_2}} \right)(t')
  \quad,
\end{eqnarray}
which is the usual non-Markovian Boltzmann equation with a collision
term for the electron-electron interaction in Born approximation.

%
%
%
%
%

\section{Diagrams}
\label{sec:dia}

The diagrams described in this section represent the differential
equations for the correlations. Therefore they contain only one
interaction vertex in contrast to ordinary Feynman graphs. In addition
there are graphic elements for the correlations, the so-called
correlation bubbles \cite{Sch}, which in our context stand for
time-dependent quantities and not for initial correlations.  Note that
there is no time or energy integration associated with a diagram as it
characterizes the temporal evolution of the correlation at time $t$.

\begin{figure}[htbp]
  \begin{center}
    \leavevmode \epsfig{file=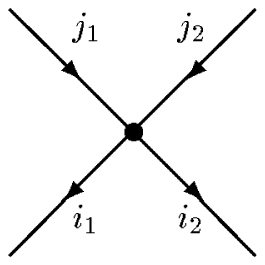}
  \end{center}
  \caption{An interaction vertex.}
  \label{fig:int_vertex}
\end{figure}

\begin{figure}[htbp]
  \begin{center}
    \leavevmode \epsfig{file=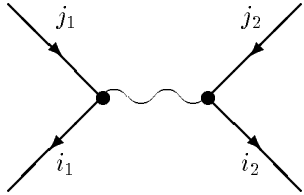}
  \end{center}
  \caption{An interaction vertex with paired creation and annihilation
    operators.}
  \label{fig:int_vertex_pair}
\end{figure}

\begin{figure}[htbp]
  \begin{center}
    \leavevmode \epsfig{file=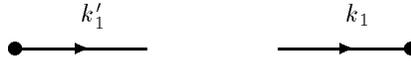}
  \end{center}
  \caption{External vertices.}
  \label{fig:ext_vertex}
\end{figure}

\begin{figure}[htbp]
  \begin{center}
    \leavevmode \epsfig{file=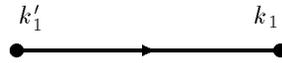}
  \end{center}
  \caption{A contraction.}
  \label{fig:contraction}
\end{figure}

\begin{figure}[htbp]
  \begin{center}
    \leavevmode \epsfig{file=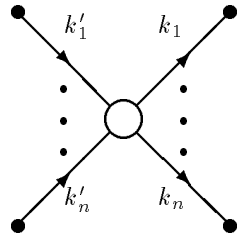}
  \end{center}
  \caption{A correlation bubble representing $\langle
    \psi_{k_1}\cdots\psi_{k_n} \psi_{k'_n}^{\dag}
    \cdots\psi_{k'_1}^{\dag} \rangle_t^c$.}
  \label{fig:corr_bubble}
\end{figure}

\begin{figure}[htbp]
  \begin{center}
    \leavevmode \epsfig{file=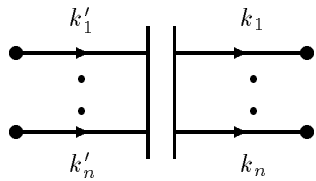}
  \end{center}
  \caption{A correlation bubble with paired creation and annihilation
    operators.}
  \label{fig:corr_bubble_pair}
\end{figure}

The diagrams are derived for our example, but they are easily extended
for systems with bosons and fermions, for systems with three-particle
interactions etc. The main elements of the diagrams are the following:
\begin{description}
\item{1.~The vertex:} An \emph{internal} vertex
  (Fig.~\ref{fig:int_vertex}) represents the interaction. The matrix
  element $v_{i_1 i_2, j_1 j_2}$ is associated with it. Incoming
  (outgoing) lines characterize annihilation (creation) operators. In
  order to simplify the determination of the correct sign, we
  sometimes use another diagrammatic element in which the creation
  operator $\psi_{i_n}^{\dag}$ is paired with the annihilation
  operator $\psi_{j_n}$ ($n=1,2$) (Fig.~\ref{fig:int_vertex_pair}).

  An \emph{external} vertex (Fig.~\ref{fig:ext_vertex}) represents a
  single creation or annihilation operator.

\item{2.~The contraction:} A contraction (Fig.~\ref{fig:contraction})
  connects two vertices and refers either to the one-particle
  expectation value $\langle \psi_{k_1}\psi_{k'_1}^{\dag} \rangle_t^c
  $ or $(-\langle \psi_{k'_1}^{\dag}\psi_{k_1} \rangle_t^c )$. This
  will be explained later.
  
\item{3.~The correlation bubble:} A correlation bubble possesses 
  incoming and  outgoing lines. The total number of lines must be
  different from two, otherwise it is a contraction. For the case of a
  fermionic system only the following correlations occur:
  \begin{equation}
    \label{eq:corr_bubble}
    \langle \psi_{k_1}\cdots\psi_{k_n} \psi_{k'_n}^{\dag}
    \cdots\psi_{k'_1}^{\dag} \rangle_t^c \quad,\, n\geq 2 \quad.
  \end{equation}
  The corresponding correlation bubble is shown in
  Fig.~\ref{fig:corr_bubble}. In order to determine the sign, it is
  better to pair the fermionic creation and annihilation operators in
  such a way that $\psi_{k_i}$ and $\psi_{k'_i}^{\dag}$ constitute a
  pair ($i=1,\ldots ,n$), which is graphically expressed as shown in
  Fig.~\ref{fig:corr_bubble_pair}.
\end{description}

We begin by describing the diagrammatic rules for the part of the time
derivative due to the interaction of the $n$-particle expectation
values $\langle \psi_{k_1}\cdots\psi_{k_n}
\psi_{k'_n}^{\dag}\cdots\psi_{k'_1}^{\dag} \rangle_t$. The
corresponding diagram consists of $2n$ external vertices and
\emph{one} interaction vertex. These vertices are connected in all
possible ways by contractions and correlation bubbles. This
corresponds to the cluster expansion of the commutator $\langle
[\psi_{k_1}\cdots\psi_{k_n}\psi_{k'_n}^{\dag}
\cdots\psi_{k'_1}^{\dag}, V] \rangle_t$. The following rules apply for
the prefactor and the sign:
\begin{description}
\item 1.~The prefactor: Usually the factor $1/{2!} \cdot 1/{2!}$ can
  be omitted due to the symmetry properties of the interaction matrix
  elements $v$ because the graph which is obtained by a permutation of
  the creation and annihilation operators in the interaction term
  yields the same contribution. For this class of graphs only one
  ``unlabeled'' graph is written down. There is only one exception if
  the interaction vertex is connected with a correlation bubble by two
  lines in the same direction, a so-called pair of equivalent lines.
  In this case a factor of $1/{2!}$ remains as the corresponding
  permutation yields the same ``labeled'' graph. In addition, there is
  always the factor $(-i)$ of Eq.~(\ref{eq:EOM}).

\item 2.~The sign: In addition to the pairing of creation and
  annihilation operators in the interaction vertex and in the
  correlation bubbles, the external vertices are grouped into pairs of
  a creation and an annihilation operator in the same way as for the
  correlations, i.e.\ $\psi_{k_i}$ and $\psi_{k'_i}^{\dag}$
  ($i=1,\ldots,n$) form a pair. Then, the overall sign is given by
  \begin{equation}
    \label{eq:sign}
    (-1)^{n} (-1)^{L} \quad,
  \end{equation}
  where $L$ is the number of fermionic loops in the deformed diagrams
  in which the paired external vertices coincide and $(-1)^{n}$ is the
  sign of the permutation normal-ordering the external pairs.  This
  sign rule can be proved in the same way as for usual Green's
  function methods \cite{Neg}.
\end{description}

\begin{figure}[htbp]
  \begin{center}
    \leavevmode \epsfig{file=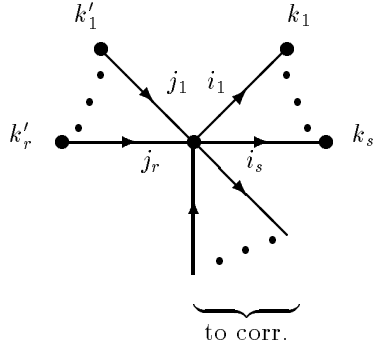}
  \end{center}
  \caption{The graph for the explanation of the Pauli factors for a
    general $n$-body vertex. There
    are $r+s$ lines connecting the interaction vertex with external
    vertices. The remaining lines go to correlation bubbles.}
  \label{fig:Pauli_fac}
\end{figure}

The description of the resulting differential equation
\begin{eqnarray}
  \label{eq:DE_uncorr}
  \lefteqn{ 
    \left[ \frac{d}{dt} +i \left( \epsilon_{k_1} +\cdots +
      \epsilon_{k_n} - \epsilon_{k'_1} -\cdots -\epsilon_{k'_n}\right)
    \right] \langle \psi_{k_1} \cdots \psi_{k_n} \psi_{k'_n}^{\dag}
    \cdots \psi_{k'_1}^{\dag} \rangle_t }
  \nonumber \\ &=& -i
  \sum_{\mathrm{diagr.}} (-1)^{n +L} \frac{1}{2^{n_e}} \sum_{i_1,i_2
    \atop j_1,j_2} v_{i_1 i_2, j_1 j_2} X_{\mathrm{diagr.}} \quad,
\end{eqnarray}
($n_e =$ number of pairs of equivalent lines, $L$ is introduced
above), is thus 
complete except for the ``collision term'' $X_{\mathrm{diagr.}}$,
which is composed of the contractions and the correlations. In the
cluster expansion every partition of $\langle \psi_{k_1} \cdots
\psi_{k_n} \psi_{k'_n}^{\dag} \cdots \psi_{k'_1}^{\dag} \cdot V
\rangle_t$ corresponds to a partition of $\langle V \cdot \psi_{k_1}
\cdots \psi_{k_n} \psi_{k'_n}^{\dag} \cdots \psi_{k'_1}^{\dag}
\rangle_t$. Since the correlations are (except for the sign)
independent of the relative order of the creation and annihilation
operators, the two terms differ only in the contributions of the
contractions. Therefore the collision term $X_{\mathrm{diagr.}}$ is
the product of the following contributions:
\begin{enumerate}
\item all correlations;
\item all contractions which start and end at the interaction vertex
  in the normal-ordered form ($-\langle \psi_{i}^{\dag}\psi_{j}
  \rangle_t^c $) as the interaction term was supposed to be
  normal-ordered;
\item the contractions between external vertices in anti-normal order
  $\langle \psi_{k_i} \psi_{k'_j}^{\dag} \rangle_t^c$ as they appear
  in this order on the lhs of the differential equation;
\item the following contribution of the remaining contractions between
  the interaction vertex and the external vertices, when these lines
  are labeled as in Fig.~\ref{fig:Pauli_fac}
  \begin{eqnarray}
    \label{eq:Pauli_fac}
    \left( \langle \psi_{k_1}\psi_{i_1}^{\dag} \rangle_t^c \cdots \langle
    \psi_{k_s}\psi_{i_s}^{\dag} \rangle_t^c (- \langle
    \psi_{k'_1}^{\dag}\psi_{j_1} \rangle_t^c ) \cdots (-\langle
    \psi_{k'_r}^{\dag}\psi_{j_r} \rangle_t^c ) \right. &&
    \\ \nonumber{}
    \left.
      - (-\langle \psi_{i_1}^{\dag}\psi_{k_1} \rangle_t^c ) \cdots
      (-\langle \psi_{i_s}^{\dag}\psi_{k_s} \rangle_t^c ) \langle
      \psi_{j_1}\psi_{k'_1}^{\dag} \rangle_t^c \cdots \langle
      \psi_{j_r}\psi_{k'_r}^{\dag} \rangle_t^c 
    \right) && \quad.
  \end{eqnarray}
\end{enumerate}
At least one contraction between the interaction vertex and an
external vertex is necessary for the graph to yield a non-vanishing
contribution. Only in the contractions the non-trivial canonical
anti-commutation rules show up.

\begin{figure}[htbp]
  \begin{center}
    \leavevmode
    \epsfig{file=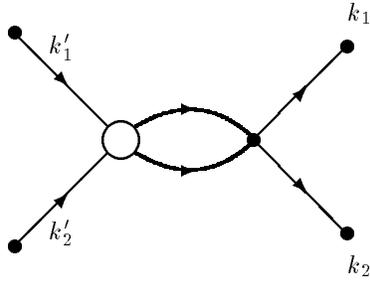}     
  \end{center}
  \caption{A graph contributing to the 4-point function.}
  \label{fig:ex1}
\end{figure}

\begin{figure}[htbp]
  \begin{center}
    \leavevmode
    \epsfig{file=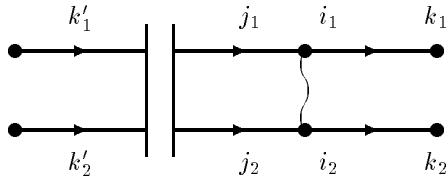}     
  \end{center}
  \caption{A labeled version of the graph in Fig.~\ref{fig:ex1} with
    paired creation and annihilation operators.}
  \label{fig:ex1_}
\end{figure}

In order to make these diagrammatic rules more transparent we give an
example. The graph shown in Fig.~\ref{fig:ex1} possesses one
pair of equivalent lines. The labeled version shown in
Fig.~\ref{fig:ex1_} has 
two loops if $\psi_{k_i}$ and $\psi_{k'_i}^{\dag}$ are paired
($i=1,2$).  Therefore it gives the following contribution to the
differential equation for $\langle
\psi_{k_1}\psi_{k_2}\psi_{k'_2}^{\dag}\psi_{k'_1}^{\dag} \rangle_t $
(cf.\ Eq.~(\ref{eq:2partEOM})):
\begin{equation}
  -i \frac{1}{2} \sum_{i_1,i_2 \atop j_1,j_2} v_{i_1 i_2, j_1 j_2}
  \left(\langle \psi_{k_1}\psi_{i_1}^{\dag} \rangle_t^c \langle
    \psi_{k_2}\psi_{i_2}^{\dag} \rangle_t^c - \langle
    \psi_{i_1}^{\dag}\psi_{k_1} \rangle_t^c \langle
    \psi_{i_2}^{\dag}\psi_{k_2} \rangle_t^c
  \right) \langle
  \psi_{j_1}\psi_{j_2}\psi_{k'_2}^{\dag}\psi_{k'_1}^{\dag}
  \rangle_t^c \quad.
\end{equation}

The diagrammatic rules for the correlations are the same as those for
the corresponding $2n$-point functions except that only connected
diagrams must be considered (linked cluster theorem). The fact that
the correlations are independent of the order of creation and
annihilation operators (except for the sign) and that the $2n$-point
functions do depend on this order does not pose any problems for the
following reason: Except for the sign, the order of creation and
annihilation operators is of importance only in contractions
connecting two external vertices. And these contractions do not appear
in connected diagrams.

\begin{figure}[htbp]
  \begin{center}
    \leavevmode
    \epsfig{file=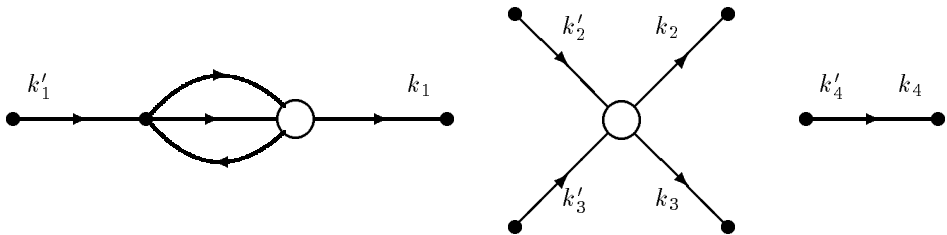}    
  \end{center}
  \caption{An unconnected graph for the 8-point function.}
  \label{fig:ex_conn}
\end{figure}

The linked cluster theorem can be proved in the following way. To each
graph for a $2n$-point function belongs a partition of the external
vertices into connected components, e.g.\ the graph shown in
Fig.~\ref{fig:ex_conn} gives the partition
$\{\psi_{k'_1}^{\dag},\psi_{k_1}\}$,
$\{\psi_{k'_2}^{\dag},\psi_{k'_3}^{\dag},\psi_{k_2},\psi_{k_3}\}$,
$\{\psi_{k'_4}^{\dag},\psi_{k_4}\}$.  The diagrammatic partition
determines exactly one term in the cluster expansion of the $2n$-point
function. In addition, there is one and only one connected component
containing the interaction vertex, in the example it is
$\{\psi_{k'_1}^{\dag},\psi_{k_1}\}$. We call this component the marked
component. The term in the cluster expansion with this marked
component determines one term in the temporal evolution of the
$2n$-point function. In order to reduce the formalism this is
explained for the example:
\begin{eqnarray}
  \label{eq:productrule}
  \lefteqn{ 
    \left[ \frac{d}{dt} +i \left( \epsilon_{k_1} +\cdots +
      \epsilon_{k_4} - \epsilon_{k'_1} -\cdots -\epsilon_{k'_4}\right)
    \right] \langle \psi_{k_1} \cdots \psi_{k_4} \psi_{k'_4}^{\dag}
    \cdots \psi_{k'_1}^{\dag} \rangle_t}
  \\ \nonumber 
  &=& \ldots + \left\{ \left[ \frac{d}{dt}
  +i(\epsilon_{k_1}-\epsilon_{k'_1}) \right] \langle
  \psi_{k_1}\psi_{k'_1}^{\dag} \rangle_t^c \right\} \langle
  \psi_{k_2}\psi_{k_3}\psi_{k'_3}^{\dag}\psi_{k'_2}^{\dag} \rangle_t^c
  \langle \psi_{k_4}\psi_{k'_4}^{\dag} \rangle_t^c + \ldots \quad,
\end{eqnarray}
i.e.\ the time derivative and the one-particle Hamiltonian act on the
marked component according to Leibniz's rule.

We will prove the linked cluster theorem by induction on the order of
the correlations, corresponding to their recursive definition in
Eq.~(\ref{eq:Cluster}). Obviously only connected diagrams contribute
to the time evolution of the contractions as at least one external
vertex must be contracted with the interaction vertex (and the other
one must be connected to the interaction vertex because of the
fermionic particle conservation). We consider now a correlation
$\langle \psi_{k_1} \cdots \psi_{k_n} \psi_{k'_n}^{\dag} \cdots
\psi_{k'_1}^{\dag} \rangle_t^c$ of order $n\geq 2$ and assume by
induction that the linked cluster theorem holds for all correlations
of order less than $n$. The set of all diagrams for the corresponding
$2n$-point function $\langle \psi_{k_1} \cdots \psi_{k_n}
\psi_{k'_n}^{\dag} \cdots \psi_{k'_1}^{\dag} \rangle_t$ is decomposed
into classes of diagrams with the same partition of the external
vertices and the same marked component. To each of the classes belongs
a term in the cluster expansion of the differential equation, as
described above. By induction, this term is described by the
corresponding class of diagrams if the marked component is of order
less than $n$. But the total of these diagrams are all unconnected
diagrams. Thus, the remaining term in the cluster expansion of the
differential equation
\begin{equation}
  \label{eq:linked_cluster}
  \left[ \frac{d}{dt} +i \left( \epsilon_{k_1} +\cdots +
    \epsilon_{k_n} - \epsilon_{k'_1} -\cdots -\epsilon_{k'_n}\right)
  \right] \langle \psi_{k_1} \cdots \psi_{k_n} \psi_{k'_n}^{\dag}
  \cdots \psi_{k'_1}^{\dag} \rangle_t^c 
\end{equation}
is represented by the connected diagrams. This completes the proof.
Of course, it is necessary to check the sign and the prefactor as
well. As far as the factor $(-i)1/{2^{n_e}}$ is concerned, this is
clear because the factor belongs to the component containing the
interaction vertex. We will not give a formalized proof for the
correctness of the sign, which should nonetheless be apparent as the
sign only reflects the necessary permutation to bring the fermionic
operators into the given order. We close with the remark that the
proof is the same for the case of a non-diagonal one-particle
Hamiltonian $h_{ij}(t)$ and that for bosons the induction starts with
the correlations of lowest order $\langle B^{(\dag)} \rangle_t^c$ and
no argument about particle-conservation is needed.

The fact that only connected diagrams contribute to the time evolution
of correlations is explained by the definition of the correlations
because we allowed for all possible partitions of the creation and
annihilation operators. Therefore, it is necessary to define the
cluster expansion for bosonic systems with non-vanishing expectation
values $\langle B^{(\dag)} \rangle_t$ as in Eq.~(\ref{eq:Cluster}) and
not as sometimes defined:
\begin{eqnarray}
  \label{ClusterAlt}
  \langle B_1 \rangle_t &=& \langle B_1 \rangle_t^c \quad, \nonumber
  \\ \langle B_1 B_2 \rangle_t &=& \langle B_1 B_2 \rangle_t^c \quad,
  \\ \langle B_1 B_2 B_3 \rangle_t &=& \langle B_1 B_2 \rangle_t^c
  \langle B_3 \rangle_t^c + \langle B_2 \rangle_t^c \langle B_1 B_3
  \rangle_t^c + \langle B_1 \rangle_t^c \langle B_2 B_3 \rangle_t^c +
  \langle B_1 B_2 B_3 \rangle_t^c \nonumber \\ \vdots \nonumber{}
\end{eqnarray}
Otherwise, also unconnected diagrams give contributions to correlations.

\begin{figure}[htbp]
  \begin{center}
    \leavevmode
    \epsfig{file=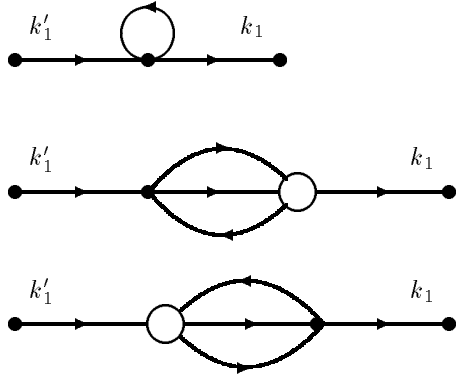}    
  \end{center}
  \caption{The diagrams for the one-particle distribution function.}
  \label{fig:1part_ex}
\end{figure}

\begin{figure}[htbp]
  \begin{center}
    \leavevmode
    \epsfig{file=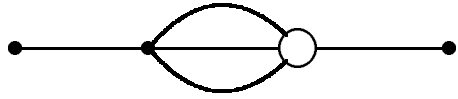}    
  \end{center}
  \caption{The diagram representing the last two diagrams in
    Fig.~\ref{fig:1part_ex}.} 
  \label{fig:1part_noarrow}
\end{figure}

\begin{figure}[htbp]
  \begin{center}
    \leavevmode
    \epsfig{file=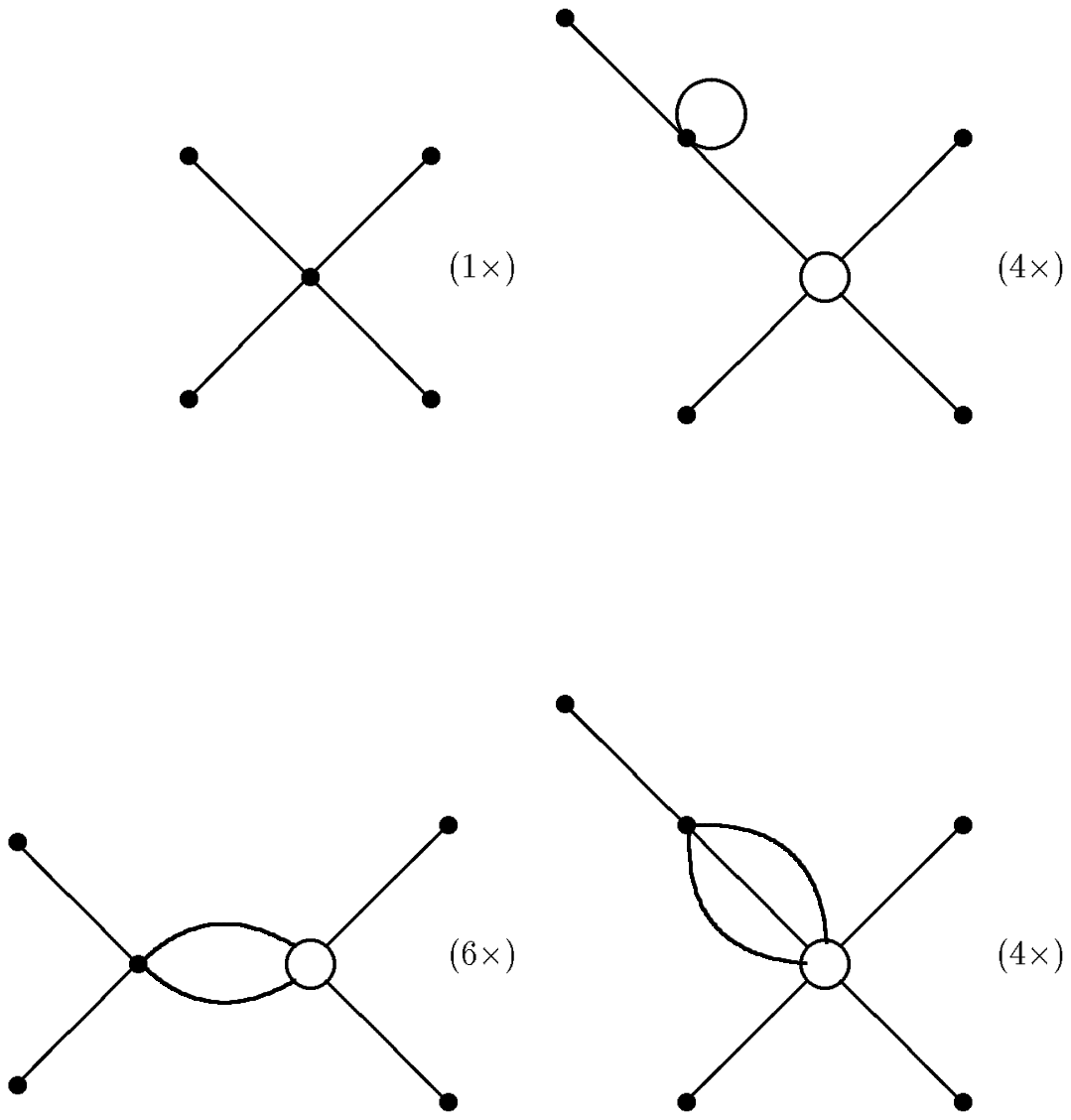}
  \end{center}
  \caption{The diagrams for the two-particle correlations.}
  \label{fig:2part_ex}
\end{figure}

\begin{figure}[htbp]
  \begin{center}
    \leavevmode
    \epsfig{file=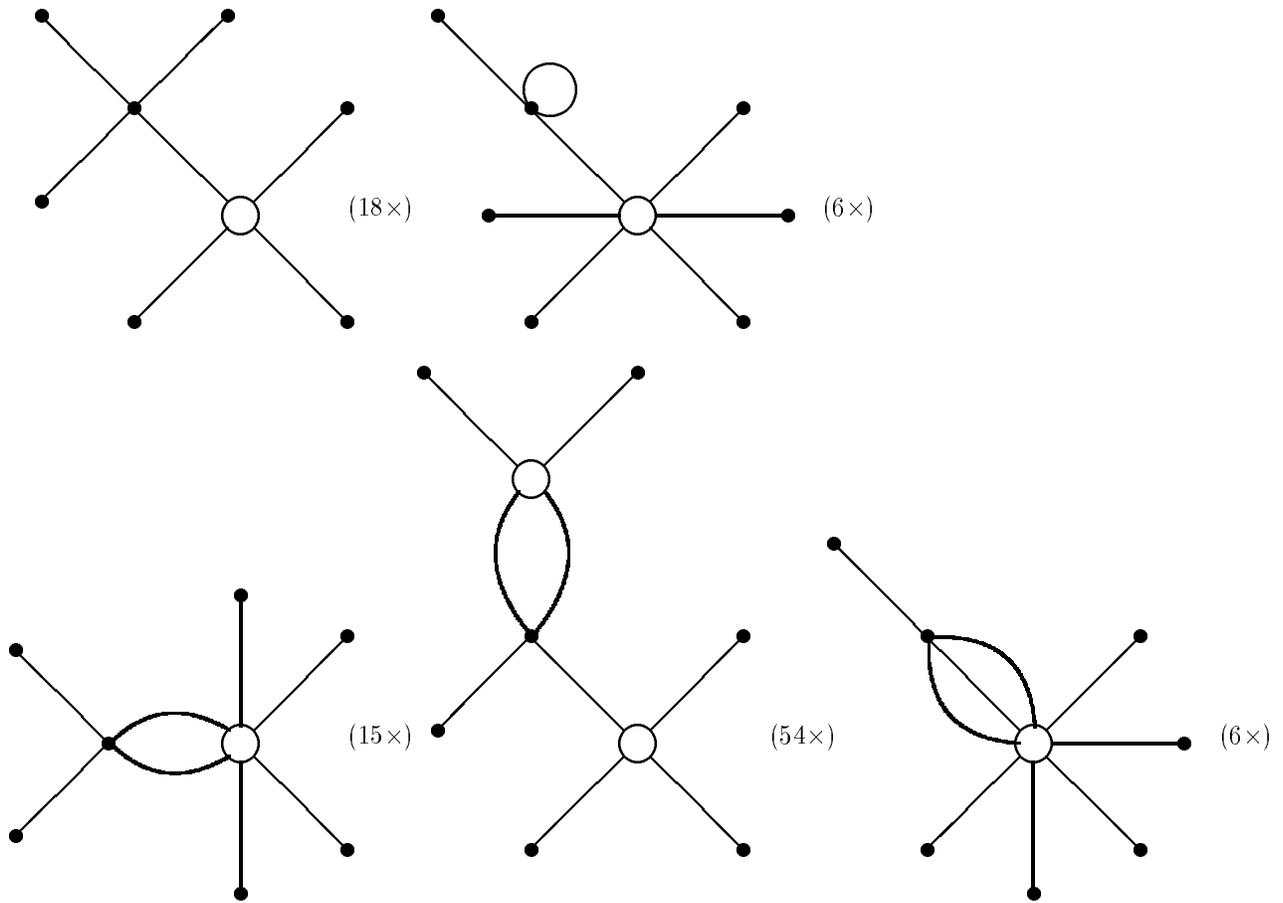}
  \end{center}
  \caption{The diagrams for the three-particle correlations.}
  \label{fig:3part_ex}
\end{figure}

We now return to the example given in section \ref{sec:EOM} and
present the diagrams for the correlations of one, two, and three
particles. The diagrams for the one-particle distribution function
representing the Hartree-Fock contribution and the ``real'' collision
term in Eq.~(\ref{eq:one-particle}) are shown in
Fig.~\ref{fig:1part_ex}. In order to greatly reduce the number of
diagrams, we draw the diagrams for higher-order correlations unlabeled
and without arrows. These diagrams are to be seen as representatives
for all the diagrams which arise by labeling the external vertices in
all possible ways and attaching arrows consistently with the
conservation of the particle number. The last two graphs in
Fig.~\ref{fig:1part_ex}, for example, are represented by the single
graph in Fig.~\ref{fig:1part_noarrow}. With this simplification there
are four types of diagrams for the two-particle correlations
(cf.~Eq.~(\ref{eq:2part-corrEOM})) shown in Fig.~\ref{fig:2part_ex}.
The numbers of resulting diagrams with external labels and arrows are
indicated in parentheses. For the three-particle correlations we show
the diagrams in Fig.~\ref{fig:3part_ex} without writing down the
corresponding equations. To cut off the hierarchy at this order the
four-particle correlations are ignored as described in section
\ref{sec:EOM}.

%
%
%
%
%

\section{Summary}

We have described the general formalism for a new method of deriving
transport equations for quantum many-body systems. It is based on the
equation-of-motion technique and a cluster expansion of many-particle
expectation values. Combining these two elements one obtains an exact
infinite hierarchy of EOM in terms of correlations which
allows a consistent, unambiguous breaking at any order in contrast to
the usual hierarchy. The transport equations consist of a set of
ordinary differential equations for the one-particle distribution
functions and many-particle correlations up to a given order.
By the initial values of these dynamical quantities information
(initial correlations) of the initial statistical operator enters the
quantum kinetics of the system. In contrast to real-time Green's
function methods our method involves only single-time quantities from
the beginning. There is no need for a generalized Kadanoff-Baym ansatz
and the separate determination of retarded/advanced Green's functions.
A diagrammatic representation of the equations of motion for the
correlations was derived. It simplifies the calculations in
the same way as Feynman diagrams do in equilibrium theory. 
The definition of the correlations by a cluster expansion leads to a
linked cluster theorem for the corresponding diagrams.

In this first part of a series of two papers we developed the general
formalism. The application of the transport equations to physical
systems is left to the second part.

\acknowledgments
The author would like to thank V.~Meden, C.~W\"ohler,
and K.~Sch\"{o}nhammer for stimulating discussions. This work was
financially supported by the Deutsche Forschungs\-gemein\-schaft (SFB 345
``Festk\"{o}rper weit weg vom Gleichgewicht'').

%
%
%
%
%
%

%
%
%
%
%
%

\end{document}